\documentclass[prl,twocolumn,nofootinbib, preprintnumbers, superscriptaddress]{revtex4}

\usepackage{amsmath,amssymb,slashed,braket}
\usepackage{graphicx}
\usepackage{epstopdf}
\usepackage{float,appendix}
\usepackage[colorlinks=true,
            linkcolor=blue,
            urlcolor=blue,
            citecolor=green,          
            bookmarks=true,
            bookmarksnumbered=true,
            breaklinks=true,
            pdfpagemode=Fullscreen,
            pdfstartview=FitBH]{hyperref}

\usepackage{esint}

\usepackage[normalem]{ulem}
\usepackage{color}

\definecolor{Orange}{cmyk}{0,0.61,0.87,0}
\definecolor{JungleGreen}{cmyk}{0.99,0,0.52,0}
\definecolor{OliveGreen}{cmyk}{0.64,0,0.95,0.40}
\definecolor{Brown}{cmyk}{0,0.81,1,0.60}
\definecolor{RoyalBlue}{cmyk}{0.71,0.53,0,0.12}
\definecolor{Gray}{cmyk}{0,0,0,0.40}
\definecolor{LightPink}{cmyk}{0.0,0.25,0,0}
\definecolor{LLightPink}{cmyk}{0.0,0.10,0,0}
\definecolor{LightBlue}{cmyk}{0.25,0,0,0}
\definecolor{LightGray}{cmyk}{0,0,0,0.2}

\usepackage{xcolor}
\definecolor{gesfpurple}{rgb}{0.47,0.19,0.42}

\definecolor{gesflanse}{rgb}{0.00,0.50,0.50}

\definecolor{gesfblue}{rgb}{0.08,0.42,0.76}

\definecolor{gesfred}{rgb}{1,0,0}

\definecolor{gesfwhite}{rgb}{1,1,1}

\definecolor{gesfblack}{rgb}{0,0,0}

\newcommand{\geqn}[1]{Eq.\,\hypersetup{linkcolor=blue}(\ref{#1})\hypersetup{linkcolor=blue}}
\newcommand{\gfig}[1]{{\hypersetup{linkcolor=violet}Fig.\,\ref{#1}\hypersetup{linkcolor=blue}}}

\graphicspath{{figs/}}

\begin{document}

\title{Proposal for a
Quantum Mechanical Test of Gravity at Millimeter Scale }

\author{Yu Cheng}
\email{chengyu@sjtu.edu.cn}
\affiliation{Tsung-Dao Lee Institute \& School of Physics and Astronomy, Shanghai Jiao Tong University, China}
\affiliation{Key Laboratory for Particle Astrophysics and Cosmology (MOE) \& Shanghai Key Laboratory for Particle Physics and Cosmology, Shanghai Jiao Tong University, Shanghai 200240, China}

\author{Jiadu Lin}
\email{jiadulin@sjtu.edu.cn}
\affiliation{Tsung-Dao Lee Institute \& School of Physics and Astronomy, Shanghai Jiao Tong University, China}
\affiliation{Key Laboratory for Particle Astrophysics and Cosmology (MOE) \& Shanghai Key Laboratory for Particle Physics and Cosmology, Shanghai Jiao Tong University, Shanghai 200240, China}

\author{Jie Sheng}
\email{shengjie04@sjtu.edu.cn}
\affiliation{Tsung-Dao Lee Institute \& School of Physics and Astronomy, Shanghai Jiao Tong University, China}
\affiliation{Key Laboratory for Particle Astrophysics and Cosmology (MOE) \& Shanghai Key Laboratory for Particle Physics and Cosmology, Shanghai Jiao Tong University, Shanghai 200240, China}

\author{Tsutomu T. Yanagida}
\email{tsutomu.tyanagida@sjtu.edu.cn}
\affiliation{
Kavli IPMU (WPI), UTIAS, University of Tokyo, Kashiwa, 277-8583, Japan}
\affiliation{Tsung-Dao Lee Institute \& School of Physics and Astronomy, Shanghai Jiao Tong University, China}

\begin{abstract}
\fontsize{10pt}{12pt}\selectfont

The experimental verification of the Newton law of gravity at small scales has been a longstanding challenge. Recently, torsion balance experiments have successfully measured gravitational force at the millimeter scale. However, testing gravity force on quantum mechanical wave function at small scales remains difficult. In this paper, we propose a novel experiment that utilizes the Josephson effect to detect the different evolution of quantum phase induced from the potential difference caused by gravity. We demonstrate that this experiment can test gravity quantum mechanically at the millimeter scale, and also has a potential to investigate the parity invariance of gravity at small scales.

\end{abstract}

\maketitle

\section{Introduction}

Gravity is the most well-known and fundamental force in the nature. %natural world. 
However, although the law of gravity has been extensively tested on large scales \cite{MICROSCOPE:2022doy,Schlamminger:2007ht,Parks:2010sk,Wagner:2012ui,li2018measurements,Xue:2020spa,Peters2001,Tino:2020dsl,Williams:2012nc}, its validity on small scales has not been definitively established. Therefore, conducting experiments to test gravity across various scales is an imperative scientific pursuit. Additionally, 
gravity poses the most important open questions in modern physics such as its quantization.
Experimentally, quantum mechanical test of gravity is very important since it might provide us with a next step towards understanding the quantum gravity. 
Furthermore, the test of the Newton law at scales from millimeter to micrometer is well-motivated in large
extra-dimension theories \cite{Arkani-Hamed:1998jmv,Antoniadis:1990ew,Montero:2022prj}.

%The most famous table-top experiment to measure the gravitational constant $G$ based on Newton law is the torsion balance \cite{Schlamminger:2007ht,Wagner:2012ui,Parks:2010sk,li2018measurements,Xue:2020spa}.
The most renowned table-top experiments for measuring the gravitational constant $G$, according to the Newton law, is the torsion balance. It has achieved a remarkable experimental precision of $\Delta G/G \simeq 10^{-5}$ \cite{Schlamminger:2007ht,Wagner:2012ui,Parks:2010sk,li2018measurements,Xue:2020spa}. The laser cooling atomic interferometer reaches almost the same accuracy \cite{Peters2001,Rosi:2014kva} with the help of quantum techniques. These tests have mainly used macroscopic masses with mass of kilogram and at the
$\mathcal{O} (10\,$cm$)$ scale and beyond. 
However, the Newton law of the gravity had not been tested below the centimeter scale. It is quite recent that the Newton law and the equivalence principle (EP) were tested at the millimeter scale by the torsion balance experiment \cite{Westphal:2020okx}. An ultrafast photography system is also proposed to monitor the movement of objects under millimeter-scale gravity \cite{Faizal:2020tek}.
%\sout{And we have now reached at the 50 micrometer.} 
Below such a scale, even the 
presence of gravity is not certain.
Thus, it is crucial to test the gravity at the scale as small as possible. On the other hand, the quantum mechanical test of gravity, 
mainly based on the cold neutron experiments \cite{Colella:1975dq,Abele:2012dn,Nesvizhevsky:2000ba,Jenke:2014yel,Landry:2016zip,Jenke:2011zz,nesvizhevsky2002quantum}, are all reliant on the large-scale gravitational force from % of 
the Earth.

Another long-standing challenge is to detect directly the small-scale gravitational force on electrons 
%\footnote{It is known \cite{} that the Newton law of the 4-dimension gravity can be broken if we have extra dimensions. The breaking effects might be different for quarks and leptons if they have different wave functions in the extra dimensions.}. 
since the electron mass is tiny compared to those of nucleons, the verification of the equivalence principle for the electron can only be inferred indirectly from torsion balance experiments \cite{Dartora:2020etp}.

In this article, we propose the 
% a 
first quantum mechanical test of the $\mathcal{O}$(mm) scale gravity by using the superconducting Josephson junctions 
(The interplay between gravity and superconductivity is investigated in \cite{Ummarino:2020loo,DeWitt:1966yi,Modanese:1995tx,Ummarino:2019cvw,Ummarino:2017bvz}.
In particular, the effects on the JJ from the earth gravity are discussed in \cite{Ummarino:2020loo}.).
With a gravitational source nearby the junction, the Cooper pairs inside the different sides of junction feels different gravitational potential.
As a result, their phases have a different evolution,
%The two superconductors inside a junction feel different gravitational potential from a massive source and induce a phase difference across the superconductors at two sides of a junction. This 
leading to a distinct current signal generated by small-scale gravity (see \cite{Cheng:2024yrn} for the detailed discussion).
It is remarkable that our experiments can test the Newton law of the gravitational force acting on the pair of electrons (Cooper pair) \cite{Dartora:2020etp}. 
We have recently pointed out \cite{Cheng:2024yrn} that our experimental setup has a potential to test a fifth force acting on the Cooper pair. In the final section, we discuss a test of the parity conservation of gravity at small scales \cite{HariDass:1976xsb} in the present proposed experimental setup.

%Due to the small mass of electrons compared to that of baryons, 
%there is currently no experiment 
%directly detecting the small-scale gravitational effect of electrons %\gred{\footnote{??}}. 
%The verification of its equivalence principle can only be inferred 
%indirectly from torsion balance experiments \cite{Dartora:2020etp}. 
%Not to mention the detection of
%gravity-induced quantum effects of electrons. 

\section{Josephson Effect in Gravitational Potential}

A Josephson junction (JJ) 
is made of two separated superconductors and an insulator with its width $\epsilon$
in between. 
When temperature of superconductor drops below a critical point, the electron inside form Cooper pair condensation, 
entering a coherent state with a large number of particles.
This coherent state 
can be described by an order parameter, $\Psi = \sqrt{n} e^{i \phi}$ \cite{Landau:1950lwq} with 
$n$ as the Cooper pair number density.
As long as the insulator width is smaller than the coherence length of superconductor, the Cooper pairs on both sides maintain mutual coupling. Generally, 
the coherence length is of the nanometer scale, and the width of an insulator is typically $\epsilon = 1\,$nm. 
Josephson current happens if the Cooper pairs of different sides have different quantum phases, which can be used to test gravity and new physics \cite{Cheng:2024yrn}.

\begin{figure}[!t]
\centering
 \includegraphics[width=0.480
 \textwidth]{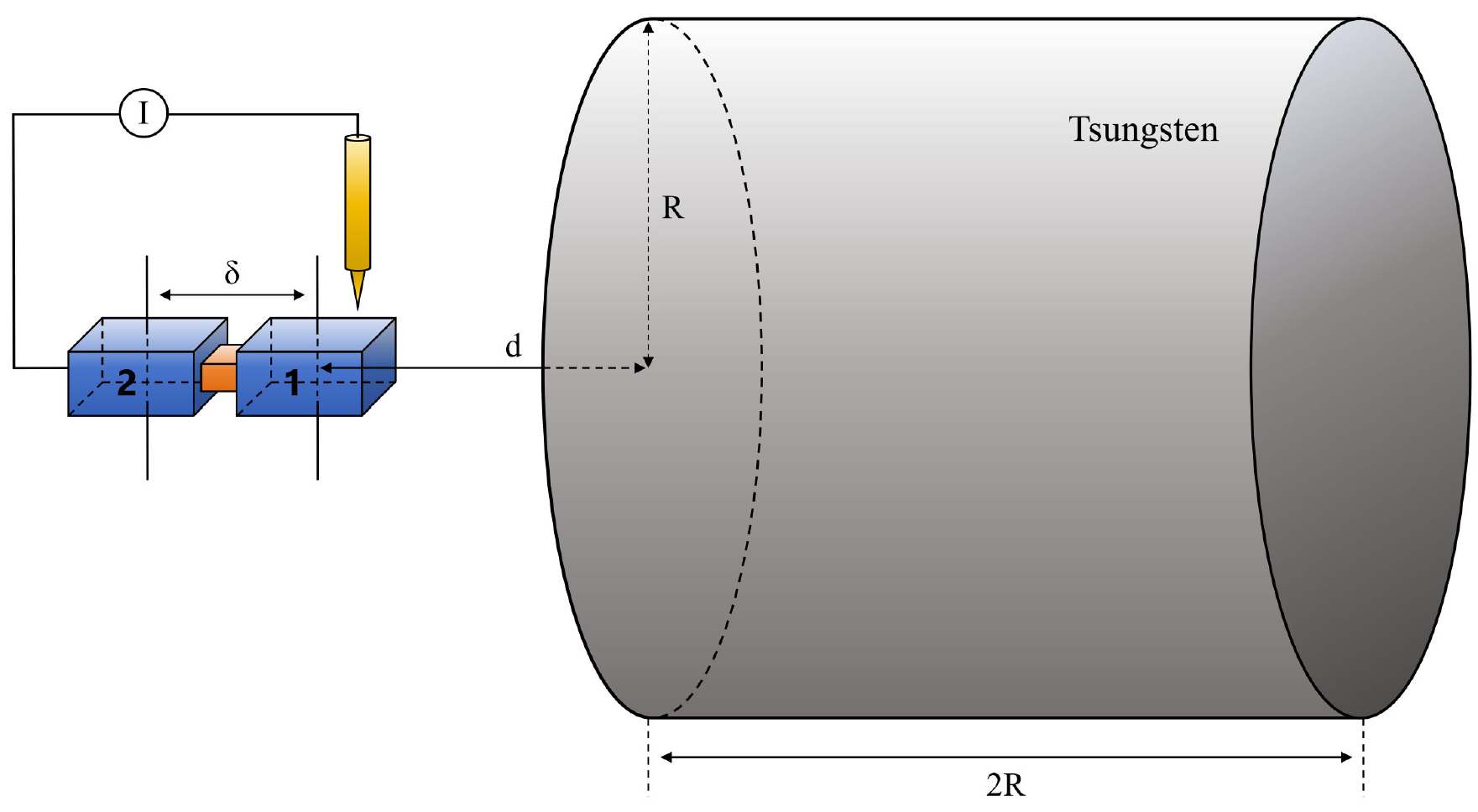}
\caption{The experimental setup for testing gravity via a Josephson junction. A tungsten cylinder with diameter and height $2R$ serves as a gravitational source and a JJ is put nearby. 
The distance from the midpoint of superconductor $1$ inside JJ and the cylinder surface is $d$. 
The average distance between two superconductors is $\delta$, which is equal to the size of the superconductor in the limit of a very thin insulator.
}
\label{fig:Design}
\end{figure}

The experimental setup is depicted in 
\gfig{fig:Design}.
The coherent states of Cooper pair in superconductor $1$ and superconductor $2$ are denoted as
$\ket{\Psi_1}$ and $\ket{\Psi_2}$, separately. Their macroscopic wave functions, 
\begin{equation}
    \Psi_1 = \sqrt{n_1} e^{i \phi_1}
\,,
\quad 
    \Psi_2 = \sqrt{n_2} e^{i \phi_2},
\label{SCwave}
\end{equation}
follow the Schrodinger equation. The number density 
$n$ is almost a constant inside superconductor and 
the phase evolves with time as $i \phi \sim i H t$.
With a gravitational source placed at a distance of $d$ on one side of JJ, 
the spatial symmetry of the entire configuration is disrupted.
As a result, the Cooper-pair state inside different
superconductors feels different gravitational potentials, and thus  
have different phases after a period of evolution. To maximum the effect
of the gravitational potential, the 
source should be made of high density material, such as tungsten
with a density $\rho = 19.5\,$g/cm$^3$. Its shape is a cylinder (Compared to a typical spherical gravitational source, the 
mass element on the surface of a cylinder is closer to the target, resulting in a larger gravitational potential.) with both diameter and height equal to $2R$. 

At the beginning, we connect the JJ to a circuit loop to consume its initial phase difference to $\Delta \phi = 0$ \cite{Cheng:2024yrn}. 
Then, we disconnect the circuit and
introduce the gravitational source by moving the cylinder to nearby the JJ. After a time period $\tau$, the source induces the two 
gravitational phases as,
\begin{align}
    \phi_1 (\tau) 
&=
    G m^* \rho \tau  
    \int_0^{2R} \int_0^R \frac{2 \pi r dr dz}{\sqrt{r^2 + (d+ z)^2}}
\\
    \phi_2 (\tau) 
&=
    G m^* \rho \tau  
    \int_0^{2R} \int_0^R \frac{2 \pi r dr dz}{\sqrt{r^2 + (d+ \delta + z)^2}}.
\end{align}
with the integration of the 
volume in cylindrical coordinate system. The phase difference becomes, $\Delta \phi (\tau) \equiv \phi_1 (\tau) - \phi_2 (\tau)$.
The separation of the two superconductor $\delta$ is the distance between the midpoints of two superconductors and $d$ is the distance between midpoint of superconductor $1$ and the surfance of cylinder (In principle, Cooper pairs at different positions within the same superconductor undergo independent phase evolutions. However, in a coherent state, all particles have the same phase, which is ensured by a small variation in the number density, as the gradient of the phase implies the generation of a current. Therefore, the phase of the coherent state should be an average of the phases of particles at different positions, effectively representing the phase at the midpoint of the superconductor.). 
%These lengths have the hierarchy
%$R \gg d \gg \delta$.
The superconductor size $a$ is much larger
than the insulator width $\epsilon$.
In such a case, the separation $\delta$ is roughly the side length $a$ of superconductor.
The $m^* = 2 m_e$ is the mass of a Cooper pair since the gravity directly acts on the wave function of a Cooper pair (Although it is very close to $2 m_e$, the exact value of Cooper pair mass is still uncertain.
Our proposal would also provide a new way to measure the gravitational mass of Cooper pairs.).

The effect of the insulator in the middle can be modeled as a constant barrier potential $V$ higher
than the kinetic energy $E$ of cooper pairs.
In this region, the wave functions can be parameterized as \cite{gross2016applied}, 
\begin{equation}
    \Psi (x)
=
C_1     \cosh{x/\xi} + 
C_2    \sinh{x/\xi},
\end{equation}
with $\xi = \sqrt{1/ 4 m_e (V - E)}$.
To match the boundary conditions \geqn{SCwave},the coefficients should be,  
\begin{equation}
\hspace{-3 mm}
C_1 =
    \frac{\sqrt{n_1} e^{i \phi_1} 
    + \sqrt{n_2} e^{i\phi_2} }{2 
    \cosh (\epsilon/2\xi) }\,,
C_2 =
    \frac{\sqrt{n_1} e^{i \phi_1} 
    - \sqrt{n_2} e^{i\phi_2} }{2 
    \sinh (\epsilon/2\xi) }.
\end{equation}
The phase difference between the boundaries gives rise to a phenomenon known as the Josephson effect \cite{Josephson:1962zz}, wherein a quantum tunneling current is generated.
The current density is generally a gauge invariant form, 
$J = - (2 e/m_e) {\rm Re} [\Psi^* (i \nabla - e\mathbf{A}) \Psi]$.
In the absence of extra voltage and magnetic field as explained in the next section, one can fix the gauge to $\mathbf{A} = 0$ without changing physics.
The current is 
\cite{Terry:1901}, 
\begin{equation}
    J (\tau) = \frac{e \sqrt{n_1 n_2}}{m_e \epsilon} \sin [\Delta \phi (\tau)]\,.
\end{equation}
The number density of Cooper pairs is related to the London penetration length $\lambda_L$ of the superconductor material, as 
$n = m_e / (e^2 \lambda_L^2)$ \cite{London:1935}. Typically, the 
penetration length can be taken as 
$\lambda_L \simeq 50\,$nm, resulting in $n_1 = n_2 = n \simeq 1.22 \times 10^{22}$cm$^{-3}$ in a JJ. 
%Assuming the shape of superconductor to be a cubic, 
%the areal size is $a^2$, resulting a signal current at time $\tau$ as, 

We will reconnect the circuit at the time $\tau$ 
and measure a 
gravity-induced current, which is proportional to the current density,
\begin{equation}
    I (\tau) = I_c \sin [\Delta \phi (\tau)]\,,
\quad 
\text{where}
\quad I_c = \frac{\pi \Delta_s}{2 e R_N}.
\end{equation}
Here $\Delta_s \sim$ meV is the superconductive energy gap and $R_N \sim O(1)\,\Omega$ is the junction resistance in the normal state \cite{Ambegaokar:1963zz}. The critical current $I_c$ of a JJ
is typically of order mA.
In the limit $\Delta \phi \ll 1$, the current signal is simply 
$I = I_c \cdot \Delta \phi$. 
 Comparing this gravity induced current with the theoretically predicted value, one can test the Newton law of gravity.

To ensure the absence of an internal magnetic field, the dimensions of a superconductor $a$, or equivalently $\delta$, should be longer than this London penetration length. We take 
$\delta \simeq a = 3\,\mu$m as a typical value.
Notice that a larger separation $\delta$ makes the phase difference larger and thus the signal larger. 

The relationship between the predicted phase difference and the size of the gravitational source is shown in \gfig{fig:signal}. 
For illustration, we take some typical operation time
$\tau = 1\,$hour (day)  (
In our setup, the JJ circuit is not connected until the measurement, and it functions as a freely evolving pure quantum system, where quantum tunneling does not consume energy.
Therefore, its operation time $\tau$ can be long theoretically \cite{Hassani:2023mke}.)   
and distance $d = 0.1 (1)\,$mm.
It can be intuitively inferred that the signal will be positively correlated with both the operational time $\tau$ and the size of the gravitational source $R$, but negatively related to the distance $d$.

\section{Backgrounds and Projected Sensitivity}

In our proposed experiment, a JJ first evolves for a period of time under the gravitational potential to cause the phase difference $\Delta \phi$, and its tunneling current is then measured. 
Now we will discuss possible backgrounds during the experimental process.

First, there are residual electromagnetic fields between neutral materials and 
in the surrounding space. 
The electromagnetic force between two macroscopic neutral objects, known as the Casimir force \cite{Klimchitskaya:2015vra}, 
only acts on the surface of the objects and does not affect the phase of electrons inside a superconductor.
Additionally, the electromagnetic shielding of the superconductor itself ensures that the Cooper pairs inside are in an environment where $\mathbf{E}= \mathbf{B}=0$. 
The only consideration would be the external fields in the insulator region, such as the geomagnetic field. 
Therefore, the entire experimental setup should be placed inside an electromagnetic shielded container (see details in \cite{Cheng:2024yrn}).

\begin{figure}[!t]
\centering
 \includegraphics[width=0.486
 \textwidth]{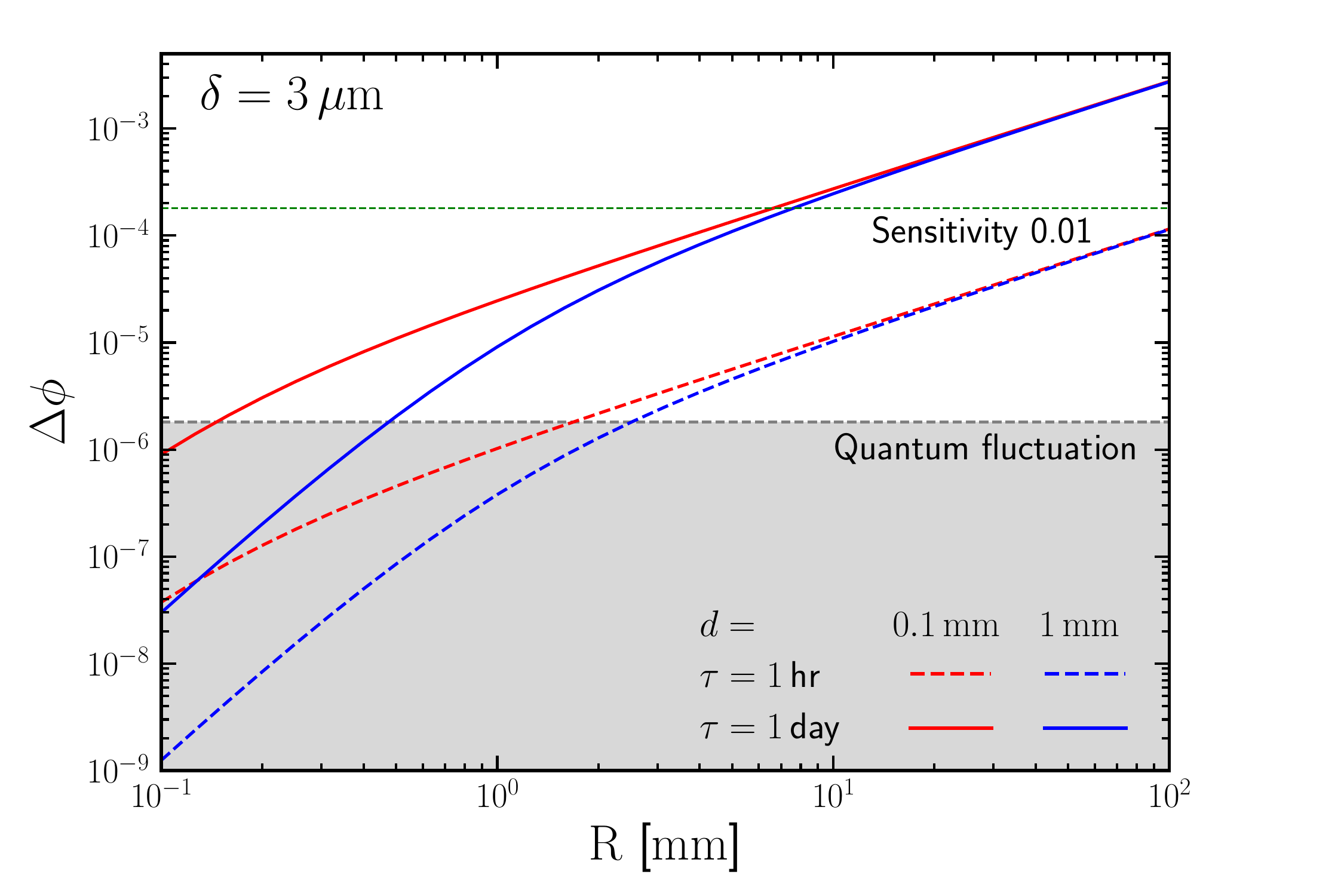}
\caption{
Phase difference $\Delta \phi$ induced by gravity potential from the massive cylinder with different radius $R$. 
The red (blue) line corresponds to a distance $d = 0.1\,$mm ($1\,$mm) while 
the dashed (solid) line corresponds to a experimental running time of one hour (day), respectively.
The grey shaded region represents for the background caused by 
the quantum fluctuation.
The horizontal green dashed line corresponds to the position where the signal is 100 times larger than the background.
}
\label{fig:signal}
\end{figure}

Second, during the evolution, the phase of a coherent state has a quantum fluctuation due to the number phase uncertainty, $ \Delta N \Delta \phi \gtrsim \hbar$ \cite{Carruthers:1965zz}. Consequently, the precise measurement of the phase 
difference is limited due to such a phase uncertainty.
For a cubic JJ with size $a \simeq \delta = 3 \mu\,$m, 
the total number of Cooper pairs is $N = 3.05 \times 10^{11}$
with a corresponding number fluctuation of $\Delta N \simeq \sqrt{N} \simeq 5.5 \times 10^{5}$.
The best achievable accuracy for phase measurement in our experiment is then $\Delta \phi \simeq 1.8 \times 10^{-6}$.

Finally, 
when we connect the circuit to do the measurement, there will be an unavoidable background thermal noise $I_T \approx e k T/ \hbar \approx 10^{-7} (T/1 K)\,$A \cite{Perez:2023tld}. 
In laboratory environments, it is possible to achieve a lower temperature of around $T = 1\,$mK, effectively reducing the thermal background to a value of $I_T = 10^{-10}\,$A. This current is smaller than that caused by quantum fluctuation.

Therefore, 
the primary background source in the experiment arises from quantum fluctuations caused by number phase uncertainty, which is shown as the grey shaded region in \gfig{fig:signal}.
Once the phase evolves beyond $\Delta \phi \gtrsim 10^{-6}$, the gravitational signal can be identified from the background.
For one day of operation,
our proposed experiment can test
the gravity 
at sub-millimeter scale.
For a shorter running time $\tau = 1\,$hour, to reach the projected detection threshold, 
the cylinder size $R$ should be over $1\,$mm.
The project sensitivity is $0.01$
once
the signal is 100 times larger than the quantum fluctuation of the phase. 
A gravitational source with size of order $\mathcal{O}(10)\,$mm can generate signal which reaches such a
sensitivity for one day of operation.
It will be the first time that the quantum mechanical test of gravity can reach the millimeter scale.

\section{Testing the Deviation of Newton Law}

The Newton law of gravity has been well tested on scales larger than centimeters. On smaller scales, gravity may exhibit different radius-scaling behaviors or be dominated by other forces. Our experiments can also test other forms of gravitational potential. Usually, the deviation from the Newton law is described by an 
extra fifth force potential. 
It can be either a Yukawa-type potential that mediates the interaction of a massive mediator particle \cite{Hoskins:1985tn,Chiaverini:2002cb,Long:2003dx,Hoyle:2004cw,Decca:2005qz,Tu:2007zz,Chen:2014oda,Perivolaropoulos:2016ucs,Tan:2020vpf}, or a confining potential that exists only at small scales \cite{Heydari-Fard:2006klr,Heydari-Fard:2007zpq}, as, 
\begin{equation}
V(r)=
\left\{
\begin{aligned}
&-G m_1 m_2 \frac{1+\alpha e^{-r / \lambda}}{r} \text{  (Yukawa-type)}\\  
&-G m_1 m_2\left[\frac{1}{r}+\alpha  r^n e^{-r / \lambda}\right]  \text{ (Confining)}.\\
\end{aligned}
\right.
\label{fifthforce}
\end{equation}

\begin{figure}[!t]
\centering
 \includegraphics[width=0.486
 \textwidth]{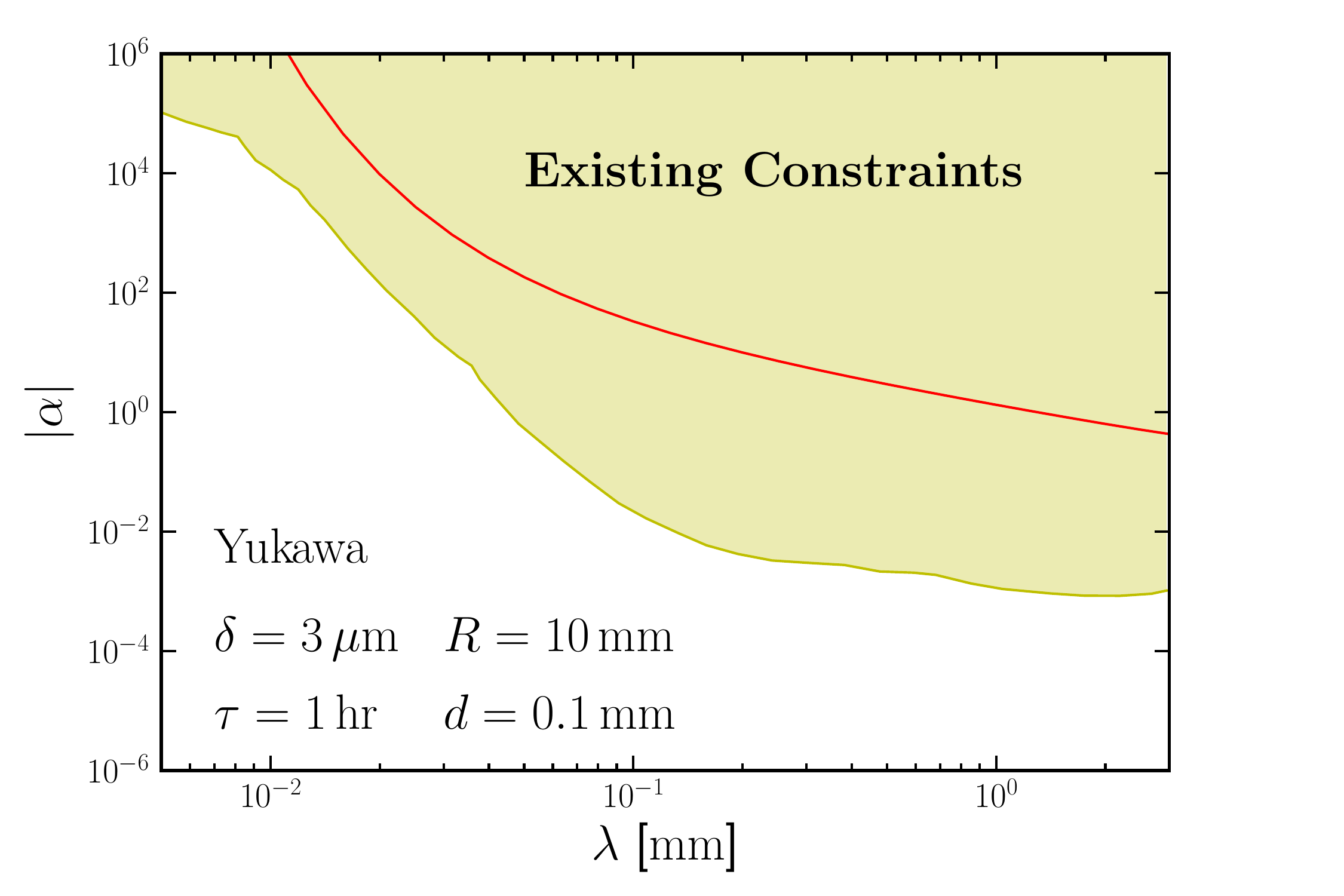}
\caption{The projected sensitivity on the $\alpha$ parameter of the Yukawa-type fifth force in our proposed setup is shown as the red line by assuming the radius of cylinder $R = 10\,$mm, distance $d = 0.1\,$mm, and running time $\tau = 1\,$hour. The yellow shaded region represents for the existing constraint on $\alpha$ \cite{Hoskins:1985tn,Chiaverini:2002cb,Long:2003dx,Hoyle:2004cw,Decca:2005qz,Tu:2007zz,Chen:2014oda,Perivolaropoulos:2016ucs,Tan:2020vpf}.
}
\label{fig:Alpha}
\end{figure}

Mainstream gravitational detection experiments, such as torsion balance, have placed strong constraints on the Yukawa-type potential \cite{Tan:2020vpf}, as indicated by the yellow region in the \gfig{fig:Alpha}. The fifth-force potential can also generate extra phase 
differences of the JJ in our setup. Once the phase difference is larger than the sensitivity $\Delta \phi = 1.8 \times 10^{-6}$, it can be detected. Assuming the parameters shown in \gfig{fig:Alpha}, the projected constraint can be achieved as the red line. In a JJ, the fifth-force potential is directly acting on the electron, whose mass is around $10^3$ times smaller than a nucleus or a neutral atom, leading to a suppression of the coupling. This is the reason why the projected sensitivity is weaker than the current constraints. Another well-motivated example is the fifth force mediated by $B-L$ gauge boson, which can serve as a dark matter candidate \cite{Cheng:2024yrn,Choi:2020kch,Okada:2020evk,Lin:2022xbu,Cheng:2023wiz,Cheng:2024vqb}. The coupling strength $\alpha$ is proportional only to the $B-L$ charge of the test bodies and thus there is a potential between a neutral material and Cooper pairs. 
Without mass suppression, the constraint on the $B-L$ fifth force in our setup can be stronger than the existing constraints as shown in out previous work \cite{Cheng:2024yrn}.

\begin{figure}[!t]
\centering
 \includegraphics[width=0.486
 \textwidth]{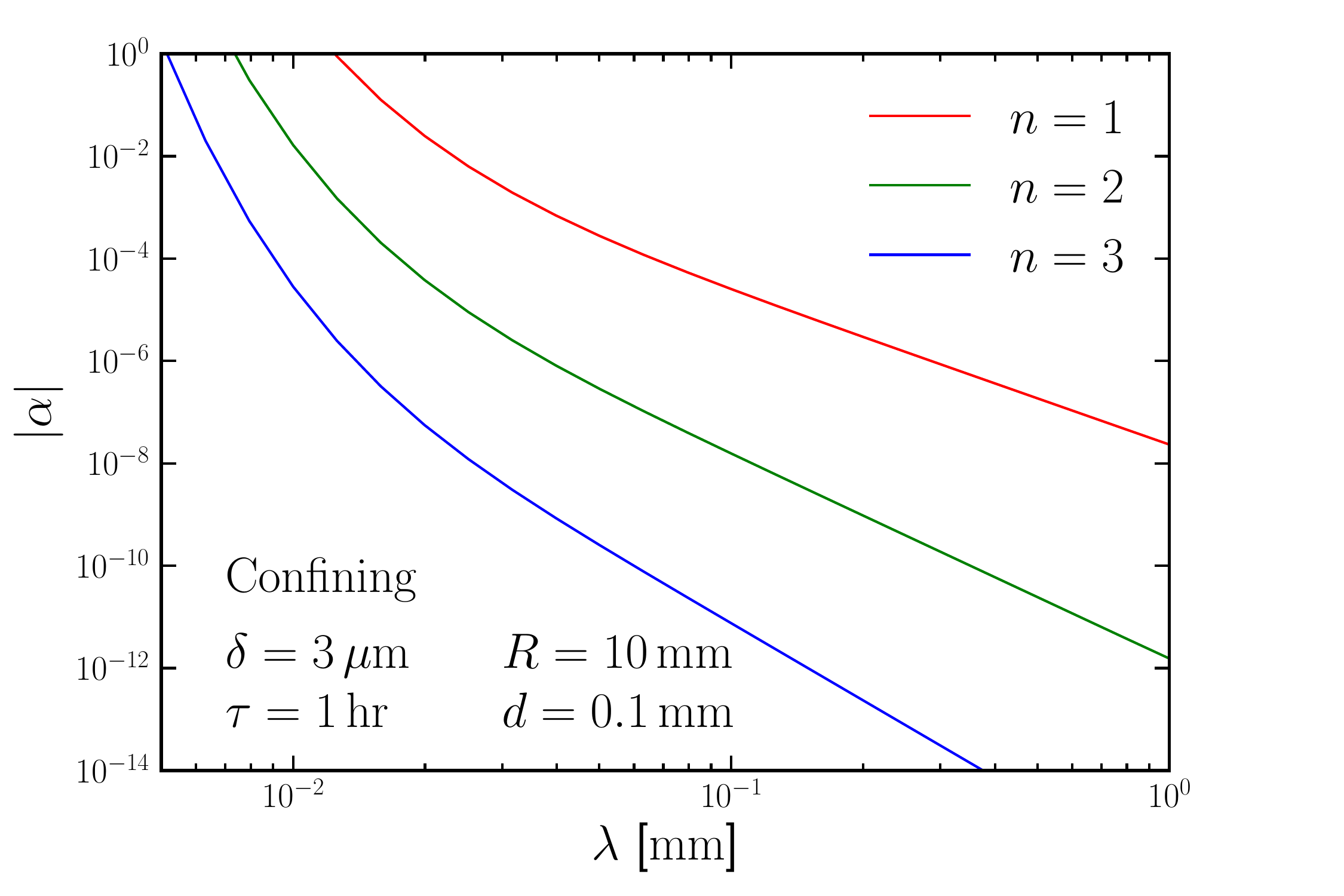}
\caption{The projected sensitivity on the $\alpha$ parameter of the confining-type gravitational force in our proposed setup. The solid lines with different colors correspond to different $r$ scalings $n$.} 
\label{fig:confine}
\end{figure}

There is relatively little exploration in experiments regarding the confining potential of gravity. Based on the same mechanism, our setup can also give projected constraints for the confining potential shown as the solid lines in \gfig{fig:confine}. Here, we take the $r$ scaling of the potential to be $n = 1, 2, 3$ as examples. At large scales, the law of gravity should revert to the Newton law, so these constraints will become very strong as lambda increases.

\section{Conclusions and Discussions}

In this paper, we propose a new 
quantum mechanical test of gravity based on Josephson effects.
By introducing a gravitational source on one side of a JJ, the wave functions of Cooper pairs in superconductor $1$ and superconductor $2$ feel different gravitational potentials. As a result, their phases of coherent states evolve independently, and the resulting phase difference generates a Josephson current which can be detected.
After the analysis of possible backgrounds, we show that our proposal can test the gravity at millimeter scale. This will be the smallest scale achievable by quantum measurement of gravity. Besides, the experimental interplay between superconductivity and gravity provides a new way to probe the undetermined Cooper pair gravitational mass.

We have shown that the 
crucial limitation of present experimental setup is given by the quantum fluctuations of the phase $\phi$. 
It might be very difficult to suppress them, since the size of superconductor is limited by the coherent length of the Cooper pairs inside. 
To enhance the signal to exceed the background more, we need a longer running time $\tau$ such more than one day. Another possibility is to use multi-JJs to enhance the Josephson currents. The operation of multi-JJs might be technically difficult, but we hope it is possible in future.

\begin{figure}[!t]
\centering
 \includegraphics[width=0.486
 \textwidth]{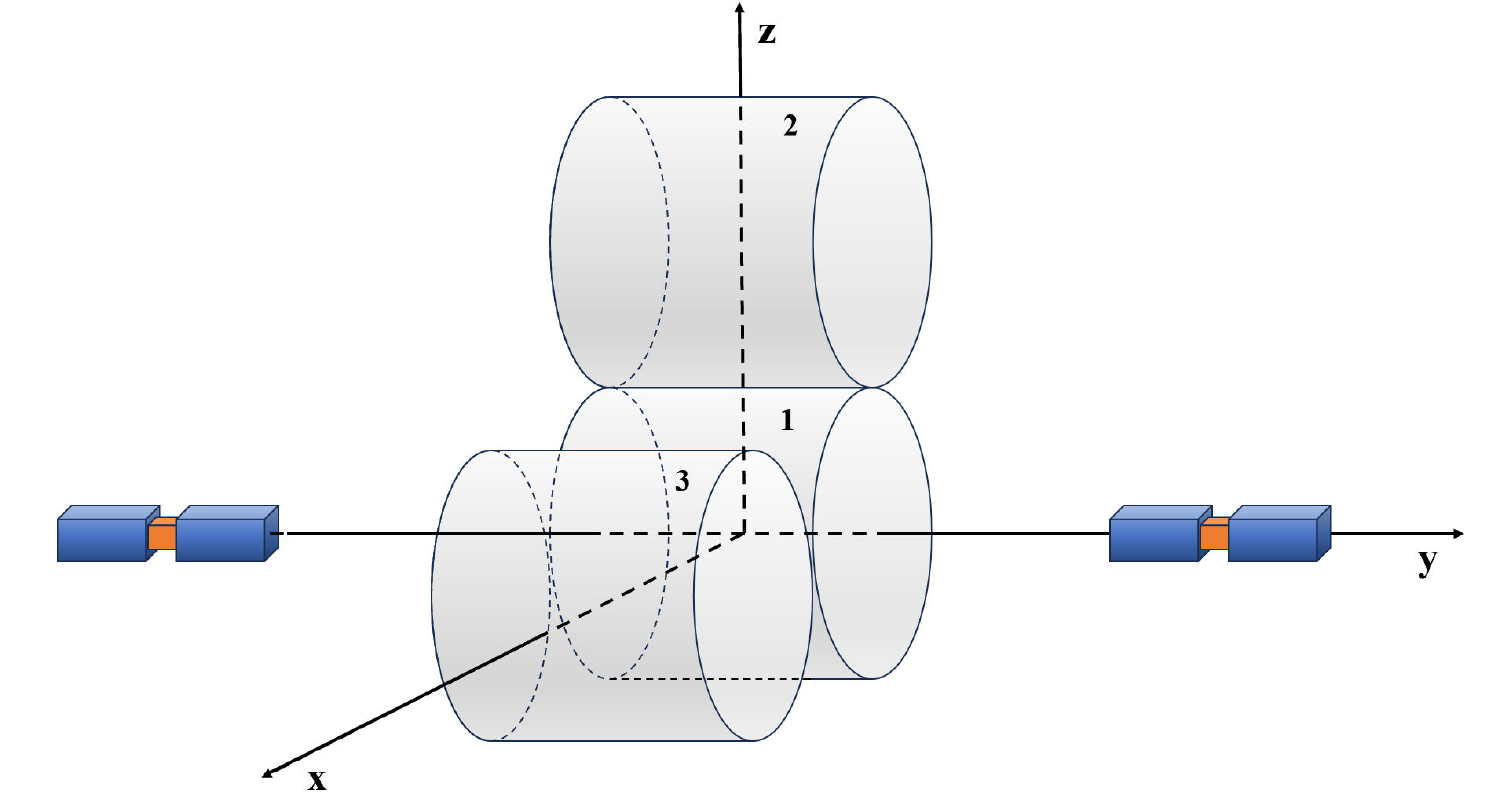}
\caption{
The experimental setup for testing parity invariance of small scale gravity using two Josephson junctions
and a gravitational source
with a chiral configuration.
Cylinder 1 is placed at at the midpoint of the line connecting of the two JJs along the y axis, while Cylinders 2 and 3 are placed above and in front of Cylinder 1 along the z and x axes respectively. Besides, Cylinder 2 and 3 should be made of different materials with different weights.
}
\label{fig:ParityDesign}
\end{figure}

We would point out that our experimental setup can provide us 
with a test of parity invariance 
(Parity violation in gravity has been extensively studied in the literature \cite{HariDass:1976xsb,Jackiw:2003pm,Alexander:2009tp} and may help explain P violation in chiral molecules and biological homochirality \cite{davankov2018biological,dorta2019homochirality}. 
The P-odd gravitational effects are tested through birefringence in gravitational wave propagation at large scales \cite{Jackiw:2003pm,Alexander:2009tp}.) of the small scale gravity. 
Usually, the parity violation of
gravitational interaction is 
parametrized as a spin-dependent Hamiltonian, which can be tested via chiral material in torsion balance \cite{Zhu:2018mrf,Dorta-Urra:2021jib}
(Similarly, the cylinder gravity source can be made of chiral material in our setup to test such a Hamiltonian.). 
Interestingly, our proposed experimental setup has a distinct advantage in detecting parity violation caused by the gravitational source with a chiral configuration, which means that, the shape of gravitational source does not satisfy the invariance under $\pi$-rotation. 
The \gfig{fig:ParityDesign} gives an example. 
We place Cylinder 1 at the origin of the coordinate axes, while Cylinder 2 and 3
made of different materials with different densities are placed along the x and z-axis directions, respectively. 
At equidistant positions on both sides of this setup along the y-axis, there are two JJs measuring the gravitational signal. 
This setup does not have a symmetry with a $\pi$-rotation around the $x-\,$and $z$-axis. 
In other words, the gravitational source has different chiralities for the two different JJs.
If the signals are different, it indicates a parity violation in gravity.

\section*{Acknowledgements}

The authors thank Tie-Sheng Yang
a lot for polishing the FIG.1. They also thank Satoshi Shirai for the discussions on the parity violation.
Yu Cheng and Jie Sheng 
would like to thank Prof. Shigeki Matsumuto for his hospitality during their
stay at Kavli IPMU where this paper was partially completed. They also thank Prof. Shao-Feng Ge for useful discussions.
This work is supported by 
the National Natural Science
Foundation of China (12175134, 12375101, 12090060, 12090064, and 12247141),
JSPS Grant-in-Aid for Scientific Research
Grants No.\,24H02244, 
the SJTU Double First Class start-up fund No.\,WF220442604,
and World Premier International Research Center
Initiative (WPI Initiative), MEXT, Japan.

\section*{Data Availability}

All data generated or analysed during this study are included in this article.

\providecommand{\href}[2]{#2}\begingroup\raggedright\endgroup

\vspace{15mm}

\end{document}